\documentclass{article}
\usepackage{spconf,amsmath,graphicx}
\usepackage{CJKutf8}
\usepackage{multirow}

\title{Monolingual Recognizers Fusion for Code-switching \\Speech Recognition}

\name{Tongtong Song$^1$, Qiang Xu$^1$, Haoyu Lu$^1$, Longbiao Wang$^{1}$, Hao Shi$^{2}$, \\ Yuqin Lin$^1$, Yanbing Yang$^1$, Jianwu Dang$^{1}$}

\address{$^1$Tianjin Key Laboratory of Cognitive Computing and Application,\\
College of Intelligence and Computing, Tianjin University, Tianjin, China\\
$^2$Graduate School of Informatics, Kyoto University, Sakyo-ku, Kyoto, Japan}
\email{\{songtongtong,longbiao\_wang\}@tju.edu.cn, shi@sap.ist.i.kyoto-u.ac.jp}
\begin{document}
% \ninept
%
\maketitle
\begin{abstract}
The bi-encoder structure has been intensively investigated in code-switching (CS) automatic speech recognition (ASR). However, most existing methods require the structures of two monolingual ASR models (MAMs) should be the same and only use the encoder of MAMs. This leads to the problem that pre-trained MAMs cannot be timely and fully used for CS ASR. In this paper, we propose a monolingual recognizers fusion method for CS ASR. It has two stages: the speech awareness (SA) stage and the language fusion (LF) stage. In the SA stage, acoustic features are mapped to two language-specific predictions by two independent MAMs. To keep the MAMs focused on their own language, we further extend the language-aware training strategy for the MAMs. In the LF stage, the BELM fuses two language-specific predictions to get the final prediction. Moreover, we propose a text simulation strategy to simplify the training process of the BELM and reduce reliance on CS data. Experiments on a Mandarin-English corpus show the efficiency of the proposed method. The mix error rate is significantly reduced on the test set after using open-source pre-trained MAMs.
\end{abstract}
\begin{keywords}
language-aware training, bilingual expert language model, text simulation, code-switching, speech recognition
\end{keywords}
\section{Introduction}
Language mixing, especially bilingual mixing, has become more prevalent with the increasing frequency of international exchanges. Bilingual mixing is also called code-switching (CS). And CS automatic speech recognition (ASR) has been studied more extensively \cite{zeng2018end,shan2019investigating,zhang2021decoupling,zhang2021end}.

Compared with large-scale monolingual data for monolingual ASR, CS ASR is limited by speech and transcripts, especially in the era of deep learning. There have been many studies on the lack of CS data problem \cite{khassanov2019constrained,winata2020meta,tseng2021mandarin}. Recently, bi-encoder structure has become a popular direction and has received much attention \cite{zhang2019towards,lu2020bi,zhou2020multi,nj2020investigation,dalmia2021transformer,yan2022joint, tian2022lae, song2022language}. In this structure, two language-specific encoders are pre-trained by the corresponding monolingual data. Then the language-specific features are extracted and then fused by shared layers to get the mixture features with bilingual information. Thus, it can take advantage of large-scale monolingual data and reduces the dependence on CS data.

However, some previous works either do not use pre-trained monolingual ASR models (MAMs) \cite{lu2020bi,dalmia2021transformer,tian2022lae}, or use MAMs with restrictions on the model structures but do not fully use the pre-trained parameters \cite{zhang2019towards,zhou2020multi,nj2020investigation,yan2022joint,song2022language}. This means that these methods cannot timely and fully utilize the pre-trained MAMs for CS ASR.

% Compared with previous works that only have language-specific encoders, some researches \cite{} show that adding language-specific loss to the corresponding language-specific encoder can improve the model's performance. \cite{yan2022joint,tian2022lae} use ``Man" and ``Eng" for masking Mandarin and English tokens in English and Mandarin parts, respectively. \cite{song2022language} uses ``unk" for both, treats another language as unknown knowledge, treats the language-specific loss as the language constraint during fine-tuning on the code-switching data, and \cite{tian2022lae} called this language-aware training strategy. For the language-aware training, tokens in another language will be masked to generate language-specific targets for calculating the corresponding language-specific loss. After the language-aware training strategy, the monolingual part can become acutely aware of another language. In addition, \cite{song2022language} proposes a language-specific characteristic assistance decoding method compatible with the training method. Moreover, it further explores that the model can recognize CS sentences without shared layers or retraining two monolingual models.

In this paper, we propose a monolingual recognizers fusion method to fuse two pre-trained MAMs for CS ASR to solve the above problem. It can be divided into the speech awareness (SA) stage and the language fusion (LF) stage. In the SA stage, two independent MAMs predict the corresponding language-specific predictions according to the acoustic features. To keep the MAMs focused on their own language, we further extend the language-aware training strategy in \cite{song2022language} to the MAMs. Meanwhile, the data from another language can also be used in the training of MAMs. In the LF stage, two language-specific predictions are fused by BELM to get the final prediction. In addition, to simplify the training process of BELM and decuple the SA stage and LF stage, we propose a text simulation strategy and make it possible for BELM to take advantage of massive text data. The experiments show that our method can directly leverage two pre-trained MAMs and can process CS ASR task even without fine-tuning two pre-trained MAMs on the CS data. And we get significant improvement by using two open-source pre-trained MAMs, which are trained by Wenetspeech \cite{zhang2022wenetspeech} and Gigaspeech \cite{chen2021gigaspeech}, respectively.

The rest of this paper is organized as follows. In Section 2, we describe our methods, including the language-aware training strategy for MAMs, the model architecture, and the text simulation strategy. In Section 3, we introduce the dataset, experimental setup, and the evaluation of our method. Finally, we conclude this paper and discuss the future work in Section 4.

\section{Monolingual Recognizers Fusion}

\subsection{Language-aware training for MAMs}
% \begin{figure*}[htbp!] 
%   \centering
%   \includegraphics[width=\textwidth]{figures/model.png}
%   \caption{The proposed two-stage code-switching speech recognition (C2) and text simulation strategy. in (b)``M*" represents the token from Mandarin vocabulary or ``unk".  ``E*" represents the token from English vocabulary or ``unk".}
%   \label{fig:model}
% \end{figure*}
% The monolingual ASR models are only trained by corresponding monolingual data. They can recognize the speech in the corresponding language very accurately. Due to its strong perception of speech, when predicting the speech parts of another language, it will predict tokens belonging to the language with similar pronunciation.

\begin{figure}[htbp!]
  \centering
  \includegraphics[width=6.5cm]{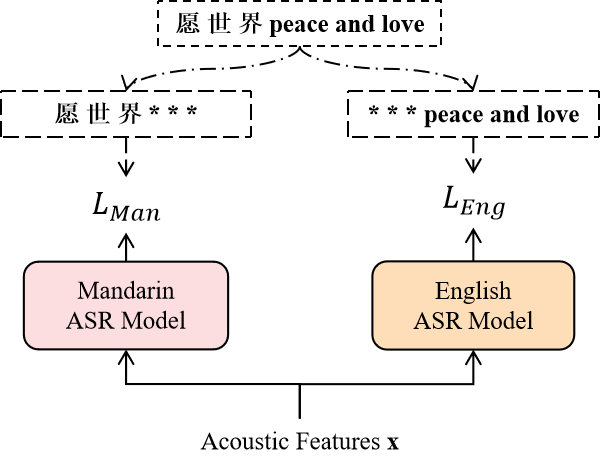}
  \vspace{-5pt}
  \caption{Language-aware training for MAMs. ``*'' represents the token ``unk''. }
  \vspace{-5pt}
  \label{fig:LA}
\end{figure}

Unlike previous works \cite{yan2022joint, tian2022lae, song2022language} only use the language-aware training strategy for the language-specific encoders of the mixture model. In our work, we further extend the language-aware training strategy to MAMs, like Fig.\ref{fig:LA}, then two MAMs are trained independently, even though they have different structures, and using the entire pre-trained parameters is more conducive to fully utilize monolingual data than only using the part of pre-trained parameters. And this makes it possible to use the data of another language for training the MAMs. 

For the language-aware training, the English and Mandarin tokens in the targets are masked by a special token to generate Mandarin-specific and English-specific targets, respectively. In this work, We use the out-of-vocabulary (OOV) token ``unk" as this special token like \cite{song2022language}. After the language-aware training, the MAMs pay more attention to the corresponding language while ignoring the more complex knowledge of another language.

\subsection{Model architecture}
\begin{figure}[htbp!]
  \centering
  \includegraphics[width=7cm]{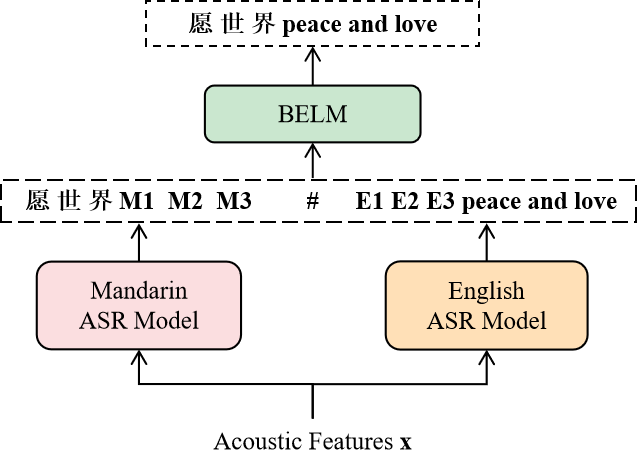}
  \caption{Model architecture. ``M*'' and ``E*'' represent the current token from Mandarin vocabulary or English vocabulary, respectively.}
  \label{fig:C2}
\end{figure}
The model architecture is shown in Fig.\ref{fig:C2}. It contains two stages: in the SA stage, two language-specific predictions $\mathbf{S}_{\rm M}$ and $\mathbf{S}_{\rm E}$ are predicted by two MAMs according to the acoustic features $\mathbf{x}$; in the LF stage, the final result $\mathbf{S}$ will be obtained by the BELM according to $\mathbf{S}_{\rm M}$ and $\mathbf{S}_{\rm E}$.
\begin{gather}
     \mathbf{S}_{\rm M}={\rm MandarinASR}(\mathbf{x}) \\
     \mathbf{S}_{\rm E}={\rm EnglishASR}(\mathbf{x}) \\
     \mathbf{S}={\rm BELM}(\mathbf{S}_{\rm M},\mathbf{S}_{\rm E})
\end{gather}
We use the attention-based encoder-decoder (AED) architecture for BELM. For the input of BELM, we concatenate two language-specific predictions and use a special label, "blank", in the middle for segmentation. During training, the decoder adopts the training strategy of teacher-forcing. During decoding, the hypotheses are decoded auto-regressively.

\subsection{Text simulation}

\begin{figure}[htbp!]
  \centering
  \includegraphics[width=7cm]{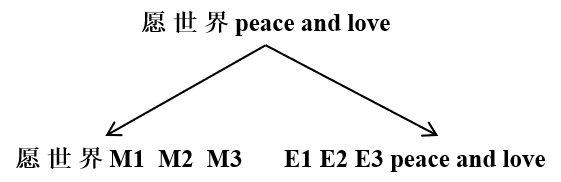}
  \vspace{-5pt}
  \caption{text simulation strategy. }
  \vspace{-5pt}
  \label{fig:TS}
\end{figure}

When training BELM, the CS speech of the training set needs to be decoded by two MAMs to generate language-specific predictions. Then they are used with the targets to train the BELM. To simplify the training process of BELM, we propose a text simulation strategy to generate the two language-specific predictions like Fig.\ref{fig:TS}.

To simulate Mandarin-specific predictions, we replace the English token in the text with a randomly selected token in the Mandarin vocabulary or "unk" with 50\% probabilities, respectively. Do the same corresponding operation to simulate English-specific predictions.

In addition, the text simulation strategy can make the SA stage and the LF stage decoupled and optimized separately so that the BELM can fully use the existing large-scale text data, thereby improving the model's performance.

\section{EXPERIMENTS}
\subsection{Dataset}
We conduct all our experiments on the ASRU 2019 Mandarin-English code-switching challenge data \cite{shi2020asru}, which consists of about 500 hours of Mandarin speech and 200 hours of intra-sentential code-switching speech for training. There are about 20 hours of speech for the development set and test set, respectively. They are all collected by smartphones in quiet rooms. The transcripts cover many common fields, including entertainment, travel, daily life, and social interaction. For English data, we use 460 hours of speech as English training set, which is the subset of Librispeech \cite{panayotov2015librispeech}.

\subsection{Experimental setup}
There are 3243 Chinese characters for the Mandarin modeling units, which are all from the transcripts of AISHELL-1 \cite{bu2017aishell} training set, and we discarded those characters whose frequency is less than 5. There are 1000 English BPE tokens for English modeling units, which are all from the transcripts of Librispeech \cite{panayotov2015librispeech} training set. In addition, there are other three special tokens ``blank"", ``unk" and ``sos/eos". And the "unk" is used for masking the tokens of another language during the language-aware training, and "blank" is used for the segmentation of two language-specific predictions to generate the input of BELM. 

For the MAMs, we extract 80-dimensional log-Mel filterbanks with a window size of 25 ms and a step size of 10 ms as the acoustic features. We normalize the features in the time dimension for each sentence. The SpecAugment \cite{park2019specaugment} is applied with 2 frequency masks ($F=10$) and 3 time masks ($T=50$) during all training stage. All MAMs are trained for 50 epochs. The maximum number of frames in one batch is 10K. The input features are sub-sampled to one quarter at the time dimension by a 2-layer convolution neural network. For the CTC-based architecture, each MAM has a 12-layer Conformer \cite{gulati2020conformer} encoder, and we use greedy search for decoding. For the AED architecture, a bi-decoder \cite{wu2021u2++} is added on the basis of the CTC-based architecture, for a 3-layer left-to-right Transformer\cite{vaswani2017attention} decoder (L2R) and a 3-layer right-to-left Transformer decoder (R2L). We use the attention rescoring decoding strategy for decoding, with CTC prefix beam search, and the beam size is 10. The weight of CTC is set to 0.3 and 0.5 during training and decoding, respectively. The weight of R2L is set to 0.3 during training and decoding.

For the BELM, it has a 4-layer Conformer encoder and 2-layer Transformer decoder. Unlike MAMs, the BELM does not have the CTC layer. We train the BELM with teacher-forcing strategy for 100 epochs, and the batch size is set to 64. The BELM decoding auto-regressively with a beam size of 10.

For the MAM and BELM, the common sets are as follows. The attention dimension is 256 with 4 heads, and the feedforward dimension is 1024. The kernel size of the convolution module in the Conformer block is 15. The models are optimized by the adam optimizer. the warmup strategy is used for the first 5 epochs with a maximum learning rate of 0.001. The dropout and label smoothing is set to 0.1 to avoid overfitting. The gradient clipping threshold is set to 5. After training, the last 5 checkpoints are averaged for decoding.

We report a mix error rate (MER) for the CS test set with the character error rate (CER) for the Mandarin part and the word error rate (WER) for the English part.

% \begin{table}[th] 
% \caption{The performance of different models. ``CS" means the model is trained by CS training set, ``All" means the model is trained by the set, which is the combination of Mandarin, English and CS training sets.}
% \label{tab:results1}
% \renewcommand\arraystretch{1}
% \centering
% \setlength\tabcolsep{3.3mm}{
%   \begin{tabular}{clccc}
%      \hline
%      \hline
%    Arc. &Model Type & CER & WER & MER \\
%    \hline
%    \multirow{4}{*}{CTC}&baseline (CS)  & 12.09 & 46.54 & 15.83 \\
%    &baseline (ALL) & 9.65 & 40.56 & 13.01 \\
%    \cline{2-5}
%    &PT &  18.11 & 79.94 & 24.83 \\
%    &FT & \textbf{9.17} & \textbf{38.29} & \textbf{12.33} \\
%     \hline
%    \multirow{6}{*}{AED}&baseline (CS) & 10.04 & 35.47 & 12.80 \\
%    &baseline (ALL) & 8.23 & \textbf{30.71} & 10.67 \\
%     \cline{2-5}
%    &PT & 16.10 & 79.16 & 22.95 \\
%    &FT & \textbf{7.8} & 31.55 & \textbf{10.40} \\
%      \cline{2-5}
%   &OSPT& 10.23 & 67.00 & 16.40 \\
%   &OSFT & \textbf{5.08} & \textbf{29.7} & \textbf{7.76} \\
%      \hline
%      \hline
%   \end{tabular}}
% \end{table}

\subsection{Evaluation of the proposed method}
We conduct experiments on two different model architectures. The experimental results of different models are shown in Table \ref{tab:results1}.
\begin{table}[htbp]
  \caption{Performance (MER\%) of different models. ``CS" means the model is trained by CS training set, ``All" means the model is trained by the set, which is the combination of Mandarin, English and CS training sets.}
  \label{tab:results1}
  \renewcommand\arraystretch{1.1}
  \centering
  \setlength\tabcolsep{6mm}{
    \begin{tabular}{clc}
       \hline
       \hline
     Architecture &Model Type &  MER \\
     \hline
     \multirow{4}{*}{CTC}&baseline (CS)  & 15.83 \\
     &baseline (ALL) & 13.01 \\
     \cline{2-3}
     &PT & 24.83 \\
     &FT & \textbf{12.33} \\
      \hline
     \multirow{6}{*}{AED}&baseline (CS) & 12.80 \\
     &baseline (ALL) & 10.67 \\
      \cline{2-3}
     &PT & 22.95 \\
     &FT & \textbf{10.40} \\
       \cline{2-3}
    &OSPT& 16.40 \\
    &OSFT & \textbf{7.76} \\
       \hline
       \hline
    \end{tabular}}
\end{table}

The baseline models adopt the single encoder. It is observed that when adding two monolingual data, the model performs better than only using the CS data for training. It indicates that monolingual data helps the model learn more monolingual information and improve the performance in their corresponding language, further improving the performance in the CS scenario.

As shown in Table \ref{tab:results1}, when using pre-trained (PT) MAMs, we can process CS ASR tasks by only training BELM. However, previous works must train the mixture ASR model with CS data. Training BELM is faster than the mixture ASR model. After fine-tuning (FT) MAMs on the CS data by the language-aware training strategy, the proposed method outperforms the corresponding baseline under both architectures.

As shown in the last two rows of Table \ref{tab:results1}, we use two open-source pre-trained MAMs (OSPT), which are trained by 1w hours of Mandarin data Wenetspeech \cite{zhang2022wenetspeech} and 1w hours of English data Gigaspeech \cite{chen2021gigaspeech}, respectively. As shown in the OSPT in Table \ref{tab:results1}, we can obtain 16.40\% MER by directly using open-source pre-trained MAMs and only training BELM. After fine-tuning open-source pre-trained MAMs on the CS training set (OSFT), we can obtain the MER of 7.76\%, which is a significant improvement over all the baselines and shows the effectiveness of the proposed method.

\subsection{Performance of BELM}
\begin{CJK}{UTF8}{gbsn}
  \begin{table*}[th!]
    \renewcommand\arraystretch{1.1}
    \caption{The predictions of two MAMs and BELM in OSPT and OSFT. We select two examples. ``U" represents the OOV token ``unk".}
    \vspace{2pt}
    \label{tab:results3}
    \centering
    \setlength\tabcolsep{0.2mm}{
    \begin{tabular}{c|c|c}
        \hline
        \hline
    Type &  Example 1 & Example 2 \\
      \hline
    GT & 非\quad 常\quad \_interpret\,er\quad \_friend\,ly\quad 以\quad 及\quad 每\quad 次\quad 口\quad 译 & 第\quad 一\quad 首\quad 歌\quad \_sleep\quad \_tight\quad 非\quad 常\quad 应\quad 景 \\
    \hline
    \multirow{3}{*}{OSPT} & 非\quad 常\quad 英\;特\;瑞\;\,特\quad 弗\;\,德\;\,利\quad  以\quad 及\quad 每\quad 次\quad 口\quad 译 & 第\quad 一\quad 首\quad 歌\quad 斯\quad 利\quad\;\;特\quad\;非\quad 常\quad 应\quad 劲 \\ 
    & \_fe\quad ta\quad \_interpret\,or\quad \_friend\,ed\,ly\quad \_it\quad\;\;\_means\quad \_se\,o\quad i & \_the\;\,\_usual\quad \_girl\, \_sleep\quad \_tight\quad \_fa\quad al\quad \_in\quad \_g \\
    & 非\quad 常\quad \_interpret\,ation\quad\ 可\qquad\; 以\quad 及\quad 每\quad 次\quad 口\quad 译 & 第\quad 一\quad 首\quad 歌\quad \_sleep\quad \_tight\quad 非\quad 常\quad 应\quad 景 \\
    \hline
      \multirow{3}{*}{OSFT} & 非\quad 常\quad\; U\quad U\quad U\qquad U\quad\;U\quad\;以\quad 及\quad 每\quad 次\quad 口\quad 译 & 第\quad 一\quad 首\quad 歌\qquad U\qquad\;\; U\quad\;\;非\quad 常\quad 应\quad 景 \\
      & \,U\quad\;U\quad \_interpret\,er\quad \_friend\,ly\quad\;\,U\quad\;U\quad\;U\quad\;U\quad\;U\quad\;U & \,U\quad\;U\quad\;U\quad\;U\quad\,\_sleep\quad \_tight\quad U\quad\;U\quad\;U\quad\;U \\
      & 非\quad 常\quad \_interpret\,er\quad \_friend\,ly\quad 以\quad 及\quad 每\quad 次\quad 口\quad 译 & 第\quad 一\quad 首\quad 歌\quad\,\_sleep\quad \_tight\quad 非\quad 常\quad 应\quad 景 \\
        \hline
        \hline
    \end{tabular}} 
  \end{table*}
\end{CJK}

\begin{table}[htbp!] 
  \renewcommand\arraystretch{1.1}
  \vspace{-5pt}
  \caption{Performance of the BELM in OSPT and OSFT. }
  \vspace{5pt}
  \label{tab:results2}
  \centering
  \setlength\tabcolsep{4mm}{
    \begin{tabular}{ccccc}
       \hline
       \hline
   Type & Model & CER & WER & MER \\
      \hline
  \multirow{3}{*}{OSPT} & Mandarin   & 19.87 & - & - \\
   &English  & - & 62.94 & - \\
   \cline{2-5}
   & BELM   & 10.23 & 67.00 & 16.40 \\
   \hline
  \multirow{3}{*}{OSFT} & Mandarin   & 4.16 & - & - \\
    &English  & - & 20.30 & - \\
    \cline{2-5}
    & BELM & 5.08 & 29.77 & 7.76 \\
       \hline
       \hline
    \end{tabular}}
    \vspace{-5pt}
\end{table}

As shown in Table \ref{tab:results2}, the BELM always brings some losses based on two language-specific predictions from MAMs. We think there are three reasons: first, the training data is limited for BELM; second, the text during training with high accuracy, but the text during testing has more noise, and then there is a large mismatch between the test and training; third, there is only text information but no speech information. Like Example 1 in the OSPT of Table \ref{tab:results3}, BELM change ``\_friend ly" to ``\begin{CJK}{UTF8}{gbsn}可(ke)\end{CJK}" according to the contextual information, because ``\begin{CJK}{UTF8}{gbsn}可(ke)以(yi)\end{CJK}" is a common word in Mandarin.

As shown in Table \ref{tab:results2}, the BELM improves the performance of Mandarin-specific predictions in OSPT. Although the input text is incorrect, it contains certain pronunciation and context information. Due to the powerful modeling ability of BELM, it can correct some tokens. Like Example 2 in the OSPT of Table \ref{tab:results3}, BELM corrects the ``\begin{CJK}{UTF8}{gbsn}劲(jin or jing)\end{CJK}" to ``\begin{CJK}{UTF8}{gbsn}景(jing)\end{CJK}", because they have similar pronunciation and  ``\begin{CJK}{UTF8}{gbsn}应(ying)景\end{CJK}" is more frequent than ``\begin{CJK}{UTF8}{gbsn}应劲\end{CJK}" in Chinese.

\subsection{Results of text simulation}

From Table \ref{tab:results4}, we can see that compared with using the predictions of two MAMs to train the BELM, the text simulation strategy has dropped significantly in the model type of PT. It is because MAM tends to recognize the tokens in another language with similar pronunciation. However, the text simulation strategy can not simulate this situation well. It leads to a big mismatch between the training and test of BELM.

However, it only has a slight decline when the model type is FT. This is because we mapped the token to ``unk" with a large probability in the text simulation strategy. At the same time, the MAMs predict more accurately after fine-tuning them on the CS data. Mapping the language-specific predictions of FT models into the correct text is less complex. 

In conclusion, we believe that this strategy has more excellent application prospects. It can simplify the training process of BELM and decouple the LF stage and SA stage, then we can take the advantage of large amounts of text data.

% \begin{table}[th]
% \caption{The performance of C2 by using text simulation strategy.}
% \label{tab:results4}
% \renewcommand\arraystretch{1}
% \centering
% \setlength\tabcolsep{3mm}{
%   \begin{tabular}{clccc}
%      \hline
%      \hline
%    Architecture &Model Type & CER & WER & MER \\
%    \hline
%    \multirow{2}{*}{CTC}&PT & 27.73  & 108.25 & 36.47 \\
%    &FT & 10.77 & 41.25 & 14.08\\
%     \hline
%    \multirow{4}{*}{AED}&PT & 25.51  & 106.61 & 34.32\\
%    &FT & 9.12 & 33.29 & 11.74 \\
%      \cline{2-5}
%   &OSPT& 27.35 & 102.74 & 35.53 \\
%   &OSFT & 5.6 & 30.62 &  8.32  \\
%      \hline
%      \hline
%   \end{tabular}}
% \end{table}
\begin{table}[htbp!]
  \caption{Performance (MER\%) of the proposed method by using text simulation strategy under different models.}
  \vspace{2pt}
  \label{tab:results4}
  \renewcommand\arraystretch{1.1}
  \centering
  \setlength\tabcolsep{6mm}{
    \begin{tabular}{clc}
       \hline
       \hline
     Architecture &Model Type & MER \\
     \hline
     \multirow{2}{*}{CTC}&PT & 36.47 \\
     &FT & 14.08\\
      \hline
     \multirow{4}{*}{AED}&PT & 34.32\\
     &FT & 11.74 \\
       \cline{2-3}
    &OSPT& 35.53 \\
    &OSFT & 8.32  \\
       \hline
       \hline
    \end{tabular}}
\end{table}

\section{Conclusion and future work}

In this paper, we proposed the monolingual recognizers fusion method to timely and fully used existing two pre-trained MAMs for CS ASR tasks. The experimental results show that our method exceeds the baseline under the CTC and AED architectures. After applying the open-source pre-trained MAMs, which are trained by 1w hours Mandarin and English corpus, respectively, the MER of the CS test set improves to 7.76\%. In addition, we further propose a text simulation strategy to simplify the training process, which can decouple the LF stage from the SA stage and make it possible to use large-scale text.

For future work, we plan to try this method on a larger corpus. We will explore the feasibility of our method without using CS data. And using pre-trained language models for the LF stage to improve the performance.

\vfill\pagebreak

\ninept
\bibliographystyle{IEEEbib}
\bibliography{main}

\begin{thebibliography}{10}

\bibitem{zeng2018end}
Zhiping Zeng, Yerbolat Khassanov, Van~Tung Pham, Haihua Xu, Eng~Siong Chng, and
  Haizhou Li,
\newblock ``On the end-to-end solution to mandarin-english code-switching
  speech recognition,''
\newblock {\em arXiv preprint arXiv:1811.00241}, 2018.

\bibitem{shan2019investigating}
Changhao Shan, Chao Weng, Guangsen Wang, Dan Su, Min Luo, Dong Yu, and Lei Xie,
\newblock ``Investigating end-to-end speech recognition for mandarin-english
  code-switching,''
\newblock in {\em ICASSP 2019-2019 IEEE International Conference on Acoustics,
  Speech and Signal Processing (ICASSP)}. IEEE, 2019, pp. 6056--6060.

\bibitem{zhang2021decoupling}
Shuai Zhang, Jiangyan Yi, Zhengkun Tian, Ye~Bai, Jianhua Tao, et~al.,
\newblock ``Decoupling pronunciation and language for end-to-end code-switching
  automatic speech recognition,''
\newblock in {\em ICASSP 2021-2021 IEEE International Conference on Acoustics,
  Speech and Signal Processing (ICASSP)}. IEEE, 2021, pp. 6249--6253.

\bibitem{zhang2021end}
Shuai Zhang, Jiangyan Yi, Zhengkun Tian, Ye~Bai, Jianhua Tao, Xuefei Liu, and
  Zhengqi Wen,
\newblock ``End-to-end spelling correction conditioned on acoustic feature for
  code-switching speech recognition,''
\newblock {\em INTERSPEECH}, pp. 266--270, 2021.

\bibitem{khassanov2019constrained}
Yerbolat Khassanov, Haihua Xu, Van~Tung Pham, Zhiping Zeng, Eng~Siong Chng,
  Chongjia Ni, and Bin Ma,
\newblock ``Constrained output embeddings for end-to-end code-switching speech
  recognition with only monolingual data,''
\newblock {\em arXiv preprint arXiv:1904.03802}, 2019.

\bibitem{winata2020meta}
Genta~Indra Winata, Samuel Cahyawijaya, Zhaojiang Lin, Zihan Liu, Peng Xu, and
  Pascale Fung,
\newblock ``Meta-transfer learning for code-switched speech recognition,''
\newblock {\em arXiv preprint arXiv:2004.14228}, 2020.

\bibitem{tseng2021mandarin}
Liang-Hsuan Tseng, Yu-Kuan Fu, Heng-Jui Chang, and Hung-yi Lee,
\newblock ``Mandarin-english code-switching speech recognition with
  self-supervised speech representation models,''
\newblock {\em arXiv preprint arXiv:2110.03504}, 2021.

\bibitem{zhang2019towards}
Shiliang Zhang, Yuan Liu, Ming Lei, Bin Ma, and Lei Xie,
\newblock ``Towards language-universal mandarin-english speech recognition.,''
\newblock in {\em INTERSPEECH}, 2019, pp. 2170--2174.

\bibitem{lu2020bi}
Yizhou Lu, Mingkun Huang, Hao Li, Jiaqi Guo, and Yanmin Qian,
\newblock ``Bi-encoder transformer network for mandarin-english code-switching
  speech recognition using mixture of experts.,''
\newblock in {\em Interspeech}, 2020, pp. 4766--4770.

\bibitem{zhou2020multi}
Xinyuan Zhou, Emre Y{\i}lmaz, Yanhua Long, Yijie Li, and Haizhou Li,
\newblock ``Multi-encoder-decoder transformer for code-switching speech
  recognition,''
\newblock {\em arXiv preprint arXiv:2006.10414}, 2020.

\bibitem{nj2020investigation}
Metilda Sagaya~Mary NJ, Vishwas~M Shetty, and Srinivasan Umesh,
\newblock ``Investigation of methods to improve the recognition performance of
  tamil-english code-switched data in transformer framework,''
\newblock in {\em ICASSP 2020-2020 IEEE International Conference on Acoustics,
  Speech and Signal Processing (ICASSP)}. IEEE, 2020, pp. 7889--7893.

\bibitem{dalmia2021transformer}
Siddharth Dalmia, Yuzong Liu, Srikanth Ronanki, and Katrin Kirchhoff,
\newblock ``Transformer-transducers for code-switched speech recognition,''
\newblock in {\em ICASSP 2021-2021 IEEE International Conference on Acoustics,
  Speech and Signal Processing (ICASSP)}. IEEE, 2021, pp. 5859--5863.

\bibitem{yan2022joint}
Brian Yan, Chunlei Zhang, Meng Yu, Shi-Xiong Zhang, Siddharth Dalmia, Dan
  Berrebbi, Chao Weng, Shinji Watanabe, and Dong Yu,
\newblock ``Joint modeling of code-switched and monolingual asr via conditional
  factorization,''
\newblock in {\em ICASSP 2022-2022 IEEE International Conference on Acoustics,
  Speech and Signal Processing (ICASSP)}. IEEE, 2022, pp. 6412--6416.

\bibitem{tian2022lae}
Jinchuan Tian, Jianwei Yu, Chunlei Zhang, Chao Weng, Yuexian Zou, and Dong Yu,
\newblock ``Lae: Language-aware encoder for monolingual and multilingual asr,''
\newblock {\em arXiv preprint arXiv:2206.02093}, 2022.

\bibitem{song2022language}
Tongtong Song, Qiang Xu, Meng Ge, Longbiao Wang, Hao Shi, Yongjie Lv, Yuqin
  Lin, and Jianwu Dang,
\newblock ``Language-specific characteristic assistance for code-switching
  speech recognition,''
\newblock in {\em Proc. Interspeech 2022}, 2022, pp. 3924--3928.

\bibitem{zhang2022wenetspeech}
Binbin Zhang, Hang Lv, Pengcheng Guo, Qijie Shao, Chao Yang, Lei Xie, Xin Xu,
  Hui Bu, Xiaoyu Chen, Chenchen Zeng, et~al.,
\newblock ``Wenetspeech: A 10000+ hours multi-domain mandarin corpus for speech
  recognition,''
\newblock in {\em ICASSP 2022-2022 IEEE International Conference on Acoustics,
  Speech and Signal Processing (ICASSP)}. IEEE, 2022, pp. 6182--6186.

\bibitem{chen2021gigaspeech}
Guoguo Chen, Shuzhou Chai, Guanbo Wang, Jiayu Du, Wei-Qiang Zhang, Chao Weng,
  Dan Su, Daniel Povey, Jan Trmal, Junbo Zhang, Mingjie Jin, Sanjeev Khudanpur,
  Shinji Watanabe, Shuaijiang Zhao, Wei Zou, Xiangang Li, Xuchen Yao, Yongqing
  Wang, Yujun Wang, Zhao You, and Zhiyong Yan,
\newblock ``Gigaspeech: An evolving, multi-domain asr corpus with 10,000 hours
  of transcribed audio,''
\newblock in {\em Proc. Interspeech 2021}, 2021.

\bibitem{shi2020asru}
Xian Shi, Qiangze Feng, and Lei Xie,
\newblock ``The asru 2019 mandarin-english code-switching speech recognition
  challenge: Open datasets, tracks, methods and results,''
\newblock {\em arXiv preprint arXiv:2007.05916}, 2020.

\bibitem{panayotov2015librispeech}
Vassil Panayotov, Guoguo Chen, Daniel Povey, and Sanjeev Khudanpur,
\newblock ``Librispeech: an asr corpus based on public domain audio books,''
\newblock in {\em 2015 IEEE International Conference on Acoustics, Speech and
  Signal Processing (ICASSP)}. IEEE, 2015, pp. 5206--5210.

\bibitem{bu2017aishell}
Hui Bu, Jiayu Du, Xingyu Na, Bengu Wu, and Hao Zheng,
\newblock ``Aishell-1: An open-source mandarin speech corpus and a speech
  recognition baseline,''
\newblock in {\em 2017 20th Conference of the Oriental Chapter of the
  International Coordinating Committee on Speech Databases and Speech I/O
  Systems and Assessment (O-COCOSDA)}. IEEE, 2017, pp. 1--5.

\bibitem{park2019specaugment}
Daniel~S Park, William Chan, Yu~Zhang, Chung-Cheng Chiu, Barret Zoph, Ekin~D
  Cubuk, and Quoc~V Le,
\newblock ``Specaugment: A simple data augmentation method for automatic speech
  recognition,''
\newblock {\em arXiv preprint arXiv:1904.08779}, 2019.

\bibitem{gulati2020conformer}
Anmol Gulati, James Qin, Chung-Cheng Chiu, Niki Parmar, Yu~Zhang, Jiahui Yu,
  Wei Han, Shibo Wang, Zhengdong Zhang, Yonghui Wu, et~al.,
\newblock ``Conformer: Convolution-augmented transformer for speech
  recognition,''
\newblock {\em arXiv preprint arXiv:2005.08100}, 2020.

\bibitem{wu2021u2++}
Di~Wu, Binbin Zhang, Chao Yang, Zhendong Peng, Wenjing Xia, Xiaoyu Chen, and
  Xin Lei,
\newblock ``U2++: Unified two-pass bidirectional end-to-end model for speech
  recognition,''
\newblock {\em arXiv preprint arXiv:2106.05642}, 2021.

\bibitem{vaswani2017attention}
Ashish Vaswani, Noam Shazeer, Niki Parmar, Jakob Uszkoreit, Llion Jones,
  Aidan~N Gomez, {\L}ukasz Kaiser, and Illia Polosukhin,
\newblock ``Attention is all you need,''
\newblock {\em Advances in neural information processing systems}, vol. 30,
  2017.

\end{thebibliography}

\end{document}